\documentclass{article}

\usepackage[margin=3.5cm]{geometry}
\usepackage{amsmath,amssymb}
\usepackage{graphicx}
\usepackage{authblk}
\usepackage{cite}
\usepackage{hyperref}
\usepackage{titlesec}
\usepackage{setspace}
\usepackage{caption}
\usepackage{booktabs}

\usepackage{siunitx}
\usepackage{todonotes}
\usepackage{graphicx}
\usepackage{epstopdf}
\usepackage{subcaption}
\usepackage{hyperref}
\usepackage{nicefrac, xfrac}
\usepackage{isotope}

\usepackage{command_definitions}

\title{Fiber-coupled external cavity DFB-laser}

\author[1]{David C. Nak\thanks{david.nak@physnet.uni-hamburg.de}}
\author[2]{Jasper Riebesehl}
\author[1]{Andreas Hemmerich}

\affil[1]{\small Institute for Quantum Physics, Universität Hamburg, 22761 Hamburg, Germany}
\affil[2]{\small Department of Electrical and Photonics Engineering, Technical University of Denmark (DTU), DK-2800 Kgs. Lyngby, Denmark}

\date{(Dated: May 22, 2025)}

\begin{document}

\maketitle

\begin{abstract} 
Narrow-linewidth, compact and robust laser systems are of high interest, especially with prospects for portable quantum metrological applications. We realized an external cavity DFB-laser incorporating an intra-cavity EOM in a fully fiber-coupled manner. Additional active feedback on the EOM, by referencing a high finesse cavity, enabled us to reach low kHz-linewidths. We analyzed the laser spectrum by means of delayed self-heterodyne measurements to reconstruct the power spectral density. This is achieved by utilizing a data-driven power spectrum equalization approach. The presented laser setup is successfully used to create an ultracold gas by laser cooling on a narrow atomic transition of \isotope[40]{Ca} at \qty{1978}{\nano\meter}. 
\end{abstract}

\section{Introduction}
Narrow-linewidth lasers are essential for various applications such as for optical clocks and quantum metrology \cite{ye2008quantum, muniz2021cavity}, laser cooling \cite{katori1999magneto, kuwamoto1999magneto, binnewies2001doppler, hobson2020midinfrared, akatsuka2021three}, quantum computation \cite{onishchenko2019frequency,trautmann2023s} and quantum communication \cite{jainPracticalContinuousvariableQuantum2022}. A well established laser technology in the field of atomic physics are semiconductor laser diodes due to their simplicity and economical prices. 
However, laser diodes usually have free-running linewidths spanning several \qty{}{\giga\hertz} limiting the applicability for the aforementioned fields. 
One possibility to address this shortcoming is to introduce frequency-selective optical feedback on the laser diode with additional external-cavity feedback \cite{dahmani1987frequency, hemmerich1990optically}.
A common implementation of this approach are grating-stabilized external-cavity diode lasers (ECDL)\cite{ricci1995compact} where an optical grating acts as a frequency discriminator. In these setups, the laser linewidth can be reduced to a few \qty{}{\kilo\hertz}. A possible limitation in these systems is typically caused by the sensitivity to environmental perturbations, as they are often implemented using free-space optics.
Another approach is the addition of an integrated frequency selective element such as in distributed feedback (DFB) or distributed Bragg reflector (DBR) lasers. 
These devices commonly have free-running linewidths of a few hundred \qty{}{\kilo\hertz} to a few tens of \qty{}{\mega\hertz} and are more robust to perturbations.
Additional methods to further decrease the linewidth have been demonstrated using additional optical feedback, for example by a fiber Bragg grating \cite{Wei:16} or by optical feedback of a tunable free-space external cavity using a piezo actuated mirror or an intra-cavity fiber-coupled waveguide electro-optical modulator (EOM) \cite{Lin:12, Yamoah:19}. 
The latter two approaches in particular offer a cost-efficient way for linewidth narrowing as they can be realized by off-the-shelf components with off-the-shelf laser diodes. 
However, they also have drawbacks, such as reduced capability for fast continuous frequency tuning and the introduction of free-space components. These components may require maintenance and compromise the robustness of the underlying semiconductor chip.

To address the shortcomings of the approaches demonstrated in the literature and combine their benefits, we present an improved laser design based on a fiber-coupled external cavity.
We utilize a fiber-coupled EOM with a reflective coating on the output fiber pigtail to give optical feedback into a DFB laser diode.
In this setup all components of the laser system are fiber-coupled with polarization-maintaining fibers.
This eliminates the need for opto-mechanical adjustments and maintenance, since the setup procedure is simple plug-and-play.
This laser setup is stabilized to a transfer cavity with a \qty{31}{\kilo\hertz} linewidth by applying active electronic feedback on the EOM in the optical feedback path.
Similar to the system shown in \cite{Yamoah:19} the diode laser is passively running.
The lock sustains mechanical shocks applied to the laser system, for example, by metallic tools.
Depending on the temperature stability of the laboratory, the lock holds for several hours or even days.

To benchmark the performance of the laser and compare the different states we measure its phase noise. While there usually are multiple approaches to perform such a measurement, the wavelength of the laser under test (LUT) prohibits many of them. 
Since neither a stable reference laser at \SI{2}{\micro\meter} nor a second, identical copy of the LUT is available, a straightforward heterodyne beat note measurement is not feasible.
Commercially available phase noise analyzers \footnote{i.e. \textit{OEwave OE4000}, \textit{HighFinesse Linewidth Analyzer}} are not available for the required wavelength band.
Another option is the conversion of the phase fluctuations into amplitude fluctuations using a reference cavity. 
This relieves the requirement for a separate laser\cite{DIRSCHERL1992131,Yamoah:19}. 
In this approach however the discrimination between amplitude and phase noise is lost and the analysis bandwidth is limited by the linewidth of the cavity. Additionally, for accurate low-frequency analysis, construction of an ultra-stable cavity is required, which increases complexity.

One of the remaining options is a self-heterodyne setup, in which the laser serves as its own reference.
A delay arm long enough to ensure incoherence between the light in the interferometer arms is infeasible since the coherence length of the laser is very long and the attenuation of \SI{2}{\micro\meter} light in standard single-mode fiber is very high.
Therefore the setup chosen is a short delay path self-heterodyne measurement. 
A short delay arm in this setup has the benefit of reduced sensitivity to fiber-induced phase noise \cite{hilwegLimitsProspectsLongbaseline2022} and manageable attenuation in the delay arm.

Because of the short delay, the remaining coherence causes unwanted interference effects which distort the measured phase noise. To get a true measure for the linewidth of the laser, we use a digital signal processing solution to remove the interference effects. The method is based on a power spectrum equalization filtering method, which has been shown to correct interference effects in simulated phase noise spectra \cite{kantnerAccurateEvaluationSelfheterodyne2023a}. To make this approach applicable to experimental data, we extend the method with a fitting routine to extract a non-parametric model of the phase noise directly from the data.

We determined a linewidth of a few \qty{}{\kilo\hertz} for the laser with active feedback. However, there are indications that this limitation is caused by vibrational and acoustical noise affecting the measurement system, suggesting an even smaller linewidth.
The laser was built to replace an old home-build Thulium:YAG laser and is used for cooling a gas of \isotope[40]{Ca} atoms in a magneto-optical trap on a narrow transition of metastable states down to \qty{20}{\micro\kelvin} \cite{quehl2001tunable, grunert2002sub}.

\section{Laser system and measurement setup}
\label{sec:setup}
\begin{figure}[ht!]
\includegraphics[width=\textwidth]{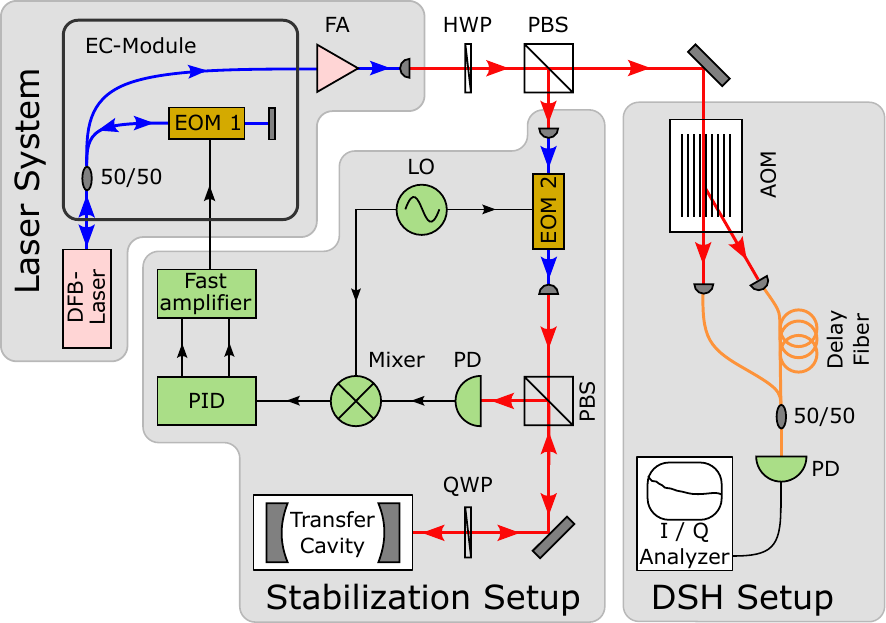}
\caption{Simplified schematic of the laser and measurement setup. 
The setup is divided into the laser system, consisting of an amplified DFB-laser with an external cavity, the stabilization setup, utilizing the PDH method for stabilization onto a transfer cavity, and a delayed self-heterodyne measurement setup. 
The laser system consists of three modules, a DFB-laser, an external cavity (EC) and a fiber amplifier, which are simply connected in series via APC fiber connectors. 
Blue lines indicate polarization-maintaining fibers, orange lines indicate single mode fibers, red lines represent free-space laser beams and black lines represent electrical connections. 
The abbreviations denote: electro-optical modulator (EOM), fiber amplifier (FA), half wave plate (HWP), quarter wave plate (QWP), polarizing beam splitter (PBS), local oscillator (LO) and photodetector (PD).}
\label{fig:SetupSchematic}
\end{figure}

The whole setup consists of three parts, the laser system, the stabilization setup and the delayed self-heterodyne measurement setup. The laser system is constructed in a modular manner. A simplified schematic of the setup is illustrated in \autoref{fig:SetupSchematic}.

\subsection{Laser system}
The seed laser used for our setup is a \texttt{Toptica DFB pro BFY} system which is based on a \texttt{Eblana EP1977-7-DM-B06-FM} distributed-feedback-like (DFB) laser diode in a butterfly package.
This butterfly package already incorporates a \qty{30}{\dB} optical isolator and has a \qty{0.8}{\m} long fiber pigtail. 
It outputs more than \qty{3}{\milli\watt} of power at our desired wavelength of \qty{1978}{\nano\meter}.
In order to separate the feedback light from the use-light, a custom made polarization-maintaining fiber splitter is used. However, for a broad range of wavelengths they are available off-the-shelf.
In order to reach a feedback power in the target range between \qty{-35}{\dB} and \qty{-47}{\dB}, we chose a target separation ratio of 50:50 for the fiber splitter. 
The pigtails of the fiber splitter are as short as the manufacturer could supply with \qty{0.3}{\m}. 

The feedback path of the splitter is then connected via a \texttt{Thorlabs ADAFCPM2} fiber-to-fiber connector to a \texttt{iXBlue MPX2000-LN-0.1-00-P-P-FA-FA} fiber-coupled waveguide electro-optic modulator (EOM). It has a \qty{150}{\mega\hertz} bandwidth and the \qty{10}{\kilo\ohm} termination of this fiber EOM enables efficient DC driving of the voltage. We shortened each pigtail of the fiber EOM from \qty{1.5}{\m} to approximately \qty{0.15}{\m}. 
Additionally, the output ferule of the fiber EOM was grinded down and polished to an output facet angle of \ang{0} and coated with alternating layers of zinc sulphide and magnesium fluoride by atomic vapor deposition to form a dielectric mirror. 
Alternatively, off-the-shelf fiber retroreflectors should work just as well.
A measurement of the total round trip losses through the splitter and fiber EOM gave a total return loss of \qty{11.8}{\dB}.
Together with the manufacturer specified \qty{30}{\dB} attenuation of the integrated isolator, the relative feedback power is estimated to be at most \qty{-41.8}{\dB}. 
The fiber splitter and EOM are together housed in a \qtyproduct{0.17x0.15x0.03}{\m} aluminum box as an external cavity module.
The laser and the fiber splitter are connected by an usual APC (angled physical contact) fiber connector.
The distance from the diode package to the mirror facet of the fiber EOM amounts to about $L=\qty{1.9}{\m}$.
The according free spectral range (FSR) $\text{c}/(2nL)=\qty{53}{\mega\Hz}$, with a refractive index for the fiber of $n=\num{1.47}$, matches with our observation.

The second branch of the fiber splitter is connected, as well by APC connectors, to the input port of a commercial two-stage thulium doped fiber amplifier, which can amplify to up to \qty{1.2}{\watt}. 
Usually the amplifier is operated at about \qty{250}{\milli\watt} output power. 
The amplifier output is again split into two parts with a half-wave-plate and a polarizing beam splitter.

\subsection{Stabilization setup}
One of these two branches is used for atomic spectroscopy and stabilization on a high finesse cavity. 
This stabilization branch contains a few \qty{}{\milli\watt} of power. This beam is coupled into a second fiber EOM (\texttt{iXBlue MPZ2000-LN-10-P-P-FA-FA}). 
It has a bandwidth of at least \qty{10}{\giga\hertz} to allow for arbitrary sideband generation. 
Focusing only on the stabilization part of this branch, the light is coupled into a high finesse transfer cavity, with a coupling efficiency of \qty{65}{\percent}. 
The retroreflected beam is picked up by a fast photo detector (\texttt{Qubig PD50-SWIR3}) using a combination of quarter-wave-plate and PBS. 
The transfer cavity setup is made up of a home-build vacuum chamber, a second, inner chamber and the resonator itself.
The vacuum chamber is made of extruded aluminum and the lid is sealed with an annealed copper gasket. 
The chamber measures \qtyproduct{0.19x0.14x0.14}{\m}. 
The two viewports and the electrical vacuum feedthrough were bought off the shelf and are made from aluminum as well.
The chamber is supported with \texttt{Sorbothane} feet on the optical table. 
The second, inner chamber is as well made from extruded aluminum and is actively temperature stabilized with a resistive heater mat and sits on four balls of FKM rubber. 
The cavity consists of a \texttt{Zerodur} spacer, a piezo-chip ring actuator and two high-reflective coated mirrors with \texttt{Infrasil} as a substrate. 
The cavity sits also on four rubber balls inside the inner chamber. 
The length of the cavity is \qty{0.1}{m}, corresponding to a FSR of \qty{1.5}{\giga\hertz}. 
The coating is designed for wavelength bands around \qty{657}{\nm} and \qty{1978}{\nm}. 
The respective linewidths are \qty{57.6}{\kilo\hertz} and \qty{30.8}{\kilo\hertz}, determined by ring-down measurements.
By applying a feedback on the piezo actuator, the cavity is actively stabilized on a highly stable reference laser, which itself is stabilized on a high finesse reference cavity \cite{Schoof:01}.
For the stabilization of the \qty{1978}{\nm} laser to the transfer cavity, a \qty{25}{\mega\hertz} modulation is applied onto the second fiber EOM.
The modulation of the EOM is done by a \texttt{Toptica PDH/DLC pro} module. 
The modulation signal is first fed through a wideband power combiner, such that a second modulation frequency can be applied, in order to set an arbitrary frequency offset from the cavity modes.
The photo detector signal is fed into a \texttt{Toptica FALC pro} PID controller with an integrated mixer. The second input of the mixer is supplied with the local oscillator frequency by the PDH module. The PID controller itself consists of two control branches. One for fast regulation with limited integration gain, usually applied directly on the laser current, and one for slow drifts with an unlimited integrator. 
As we intended to perform the complete feedback control via the intra-cavity EOM (EOM1), we customized a fast summing amplifier, based on the \texttt{Texas Instruments THS3491} operational amplifier, with a stable bandwidth of over \qty{300}{\mega\hertz} at a constant gain of \qty{15}{\dB}. 
The gain is chosen such that the output limits of the PID controller are mapped on the entire output range of the amplifier.

\subsection{Delayed self-heterodyne setup}
The second, higher power, branch of the fiber amplifier output represents the utilizable beam, which in this article is used for phase noise measurements. 
In order to do so, the light was routed through an acusto-optic deflector (AOD) (\texttt{IntraAction ACD-402AA3}) with a resonance band around \qty{40}{\mega\hertz}.
The first order deflection was then coupled into a \texttt{Corning SMF-28} single mode fiber, which we refer to as the delay fiber.
The output of the delay fiber and the zeroth order of the AOD deflection were each coupled into a separate input port of a home-made fiber combiner, made as well from \texttt{Corning SMF-28}. 
The measured attenuation in the delay fiber amounts to \qty{16.5}{\dB\per\km} at \qty{1978}{\nano\meter}, which sets a limit on the possible delay length at this wavelength. 
One of the output ports of the fiber combiner directly illuminated a \texttt{Qubig PD100-SWIR2} photodetector. 
Due to practical considerations we chose to record the beat note with an oscilloscope (\texttt{Lecroy HDO4054}) and do the analysis ex post.
To circumvent the limitation of limited recording duration per trace, imposed by the limited sample length of our oscilloscope, the beat note signal was mixed down in frequency with an analog mixer (\texttt{Mini-Circuits ZAD-1+}). As a local oscillator frequency generator a \texttt{Rhode \& Schwarz SMF 100A} was used. To suppress aliasing effects, appropriate analog low-pass filters were used between the mixer output and the input of the oscilloscope.

\section{Phase noise characterization}
\label{sec:dsp_method}

In principle, the measurement can equivalently be performed with a radio frequency spectrum analyzer instead of a real-time oscilloscope. The captured spectrum would be equivalent to $\SDelPhi$ in Eq. \eqref{eq:psd_conv_problem}, and the presented spectrum equalization method can be used identically. For completeness, however, we will detail the signal processing steps to reach $\SDelPhi$ from the time domain signal.

\subsection{I-Q demodulation}
The time domain beat note signal $y(t)$ has the form
\begin{align}
    y(t) &= \eta A(t) \cos{(\omega_{\text{AOM}}t + \Delta\phi(t))} + \epsilon_y(t) \label{eq:raw_signal} \\
    \Delta\phi(t) &= \phi(t) - \phi(t-\tau_d) \label{eq:delta_phi}
\end{align}
where $\eta$ is a detection proportionality constant, $A(t) = \sqrt{P_{CW}+\delta P_{CW}(t)}$ is the potentially noisy laser amplitude, $\omega_{\text{AOM}}$ is the beat note frequency, which is determined by the driving frequency of the AOM, $\Delta\phi(t)$ is the difference of the instantaneous phase noise $\phi(t)$ of the laser, and $\tau_d$ is the temporal delay of the long interferometer arm. 
$\epsilon_y(t)$ represents white measurement noise originating from laser shot noise and thermo-electric noise in the detector. 
Using the discrete Hilbert transform $\mathcal{H}$, the complex-valued analytic signal $Y(t) = y(t) + \mathrm{i} \mathcal{H}[y(t)]$ is estimated \cite{FELDMAN2001642}. 
This suppresses the influence of slow laser power fluctuations $\delta P_{CW}(t)$ and reduces the bias in the phase estimate and a noisy estimate of $\Delta\phi$ can be obtained from the complex argument of the analytic signal.
To characterize the phase noise $\phi(t)$, we ultimately are interested in its power spectral density (PSD) $\SPhi$, which is obtained via Fourier transformation. The subscript $\phi$ indicates the consideration of phase noise PSDs while the superscript indicates the underlying quantity.
The PSD can be formulated as
\begin{align}
    \SDelPhi = \lvert\underbrace{(1-\mathrm{e}^{\mathrm{i} 2\pi f\tau_d})}_{\equiv H(f)}
    \rvert^2 \SPhi + \SEps.
    \label{eq:psd_conv_problem}
\end{align}
The term $H(t)$ attributes for the transfer function of our short-delay self-heterodyne measurement setup, as the path delay $\tau_d$ is shorter than the coherence time of the light \cite{kantnerAccurateEvaluationSelfheterodyne2023a}.
The noise term $\SEps$ originates from the detection measurement noise of the phase $\epsilon(t)$ and is assumed to be frequency independent since the $\epsilon(t)$ is approximately normally distributed.

\subsection{Power spectrum equalization filters and data-driven approach}
\begin{figure}[h]
    \centering 
    \includegraphics[width=0.9\textwidth]{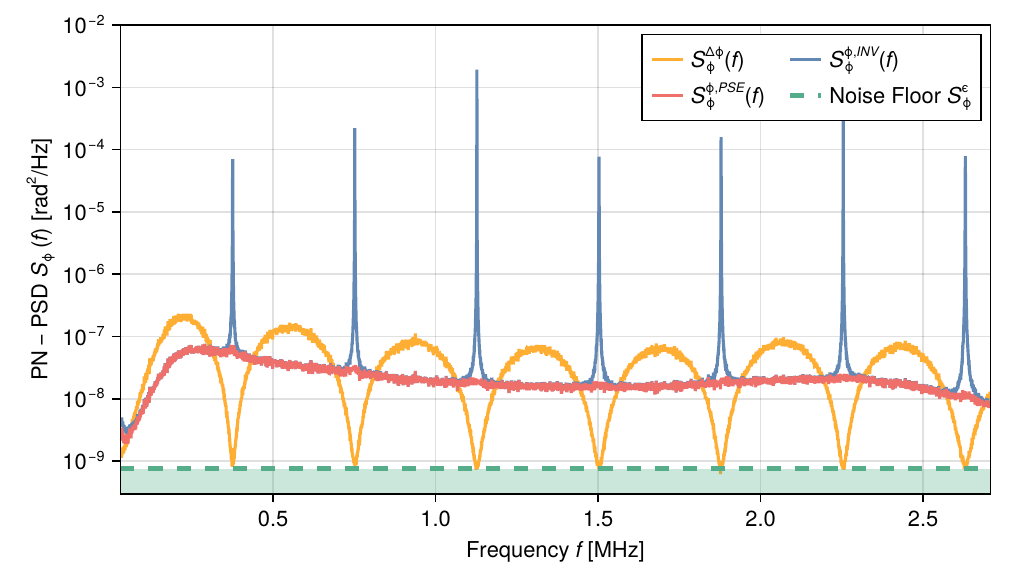}
    \caption{Visualization of the data-driven noise spectrum equalization technique. The yellow curve shows the raw phase noise PSD $\SDelPhi$ acquired with our short delay path self-heterodyne measurement setup. The blue curve shows the same data after applying the inverse transfer function of the self-heterodyne measurement setup, $\SPhiInv$. Again, the same data is visualized in the red curve showing $\SPhiPse$, using our data-driven power spectrum equalization method.}
    \label{fig:krr_method_main}
\end{figure}

A widely used approach to estimate $\SPhi$ is the multiplication of Eq. \eqref{eq:psd_conv_problem} by the inverse transfer function \cite{thorndahlthomsenFrequencyNoiseMeasurements2023, ousaidLowPhaseNoise2024, xuLaserPhaseFrequency2015}:
\begin{align}
    \SPhiInv = \SDelPhi|H(f)|^{-2} = \SPhi + \underbrace{\frac{\SEps}{2(1-\cos(2 \pi f \tau_d))} \label{eq:inv_filter}}_{\equiv \SPerturb}
\end{align}
It only requires the estimation of the delay $\tau_d$ in $H(f)$, which is found using a peak finding routine.
This method accurately estimates the true phase noise PSD when $\SPerturb$ is negligible.
However, $\SPerturb$ tends to infinity at offset frequencies that are integer multiples of $1/(2\tau_d)$, as the denominator vanishes. This problem cannot be solved by simply subtracting this term, as the quantities involved are noisy estimates.
Subsequently, $\SPhiInv$ exhibits strong artifacts at these frequencies, illustrated in Fig. \ref{fig:krr_method_main}.

An improved approach has been presented in \cite{kantnerAccurateEvaluationSelfheterodyne2023a} with power spectrum equalization (PSE) to overcome this issue. 
The method described there uses a phenomenological model describing a free running laser in order to acquire a modified transfer function and mitigate the artifacts of $\SPhiInv$.
This is not feasible in our case, as the locking mechanisms heavily modify the laser lineshape and cannot be described by a simple analytical model with few parameters.
Instead, we opt for a data-driven approach based on Kernel Ridge Regression (KRR) to use the PSE filtering approach.
KRR can be used to learn a smooth functional model for data with non-linear behavior. 
In particular, the method does not require an explicit analytic expression describing the data, making it a non-parametric method. 
The model is trained in bands of the data that are discriminated by a signal-to-noise ratio threshold.
In Fig. \ref{fig:krr_method_main}, the PSE estimate $\SPhiPse$ is visualized. 
The curve follows $\SPhiInv$ closely in the thresholded frequency regions while providing a reasonable interpolation at the poles, under the assumption that the underlying PSD is smooth. 
Other curve interpolation methods such as Gaussian process regression or spline interpolation would likely produce similar results. Our choice of method is based on its simplicity and ease of implementation. This method is discussed in more detail in \cite{riebesehl2025interferencefringemitigationshortdelay}.

\section{Results}
Using the measurement setup described in section \ref{sec:setup}, we investigate the laser phase noise properties for three different states. The phase noise is measured for the free-running laser diode without any stabilization, with just the optical feedback active and with the full locking mechanism active.
The delay arm in the interferometer was chosen to be about \SI{530}{\meter}. This choice balances the available power after the transmission through the fiber with the number of visible interference fringes in the spectrum.

\begin{figure}
    \centering
    \includegraphics[width=0.95\textwidth]{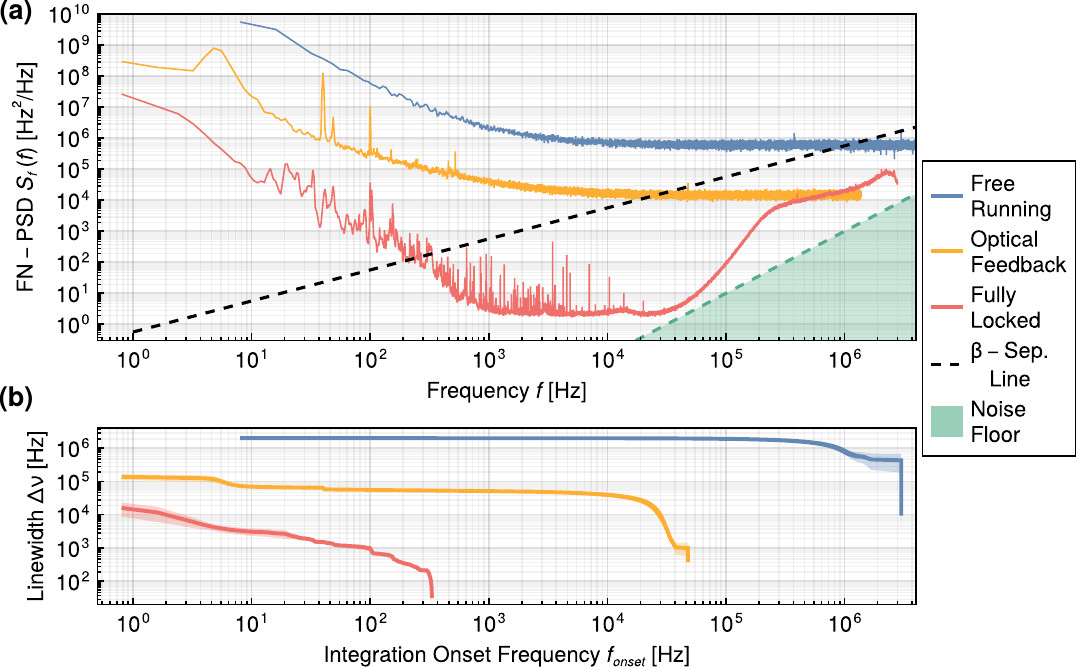}
    \caption{Laser state noise comparison. \textbf{(a)} Frequency noise power spectral densities of the laser in its different operational states after the power spectrum equalization. \textbf{(b)} Linewidth as calculated via the $\beta$ separation line method. Shaded areas are $\pm 2\sigma$ confidence intervals.}
    \label{fig:results}
\end{figure}

In Fig. \ref{fig:results}a, the frequency noise PSDs of the laser in its three different operational states are compared. 
For each state, many time domain traces are recorded and individually processed using the DSP steps detailed in Sec. \ref{sec:dsp_method}. 
For each trace, a KRR model is fitted such that the PSE filter can be calculated and applied. 
The equalized spectra of all traces per laser state are averaged. 
Finally, the phase noise PSD is converted to frequency noise PSD for convenience and visibility.
None of the interference fringes are observable in the equalized PSDs, indicating that the presented signal-processing method is robust and can effectively compensate for the interference effect.
In Fig. \ref{fig:results}b, the linewidths $\Delta\nu$ of the laser states are compared. They are calculated from the frequency noise PSDs using the $\beta$ separation line method \cite{didomenicoSimpleApproachRelation2010}. 
Since the value of $\Delta\nu$ depends on the lowest frequency considered in the integration, it is calculated for all available onset frequencies.

The free-running diode has a linewidth of about \SI{2}{\mega\hertz}. 
In this comparison, the capability of the locking mechanisms becomes clear: Optical feedback reduces the linewidth by more than an order of magnitude. 
Adding the active feedback reduces the linewidth by more than an order of magnitude again. 
The fully locked laser has a linewidth of few \si{\kilo\hertz}, dependent on integration time. 
For integration times of \SI{10}{\milli\second} or less, the linewidth is below \SI{1}{\kilo\hertz}.

\subsection{Measurement and limitations}
The data shown above were measured with the spun-up \SI{530}{\meter} delay fiber sitting in the basement below the lab. 
Still, in total about \SI{8}{\meter} of fiber had to be routed through the lab to the conduit leading into the basement. 
Measurements with the delay fiber sitting in the lab had significantly higher PSDs for frequencies up to \SI{7}{\kilo\hertz} as to sitting in the basement. 
The most obvious difference between the two locations was the soundscape which was much noisier in the lab. 
This suggests that acoustics and vibrations picked up by the fiber might be limiting in determining the actual PSD for low frequencies. 
A second delay fiber with a length of \qty{2.5}{\km} was tested as well, but no significant differences in the measurement results could be observed.
For frequencies above \SI{1}{\kilo\hertz} the spectra are reproducible over months with the fiber in the basement. 
For lower frequencies, however, variation in the spectra could be seen from day to day.
This finding and the fact that the spectrum follows a $\sim\sfrac{1}{f^2}$ behavior at low frequencies suggest that the measurement is limited by acoustic vibrations picked up by the delay fiber.
To test this hypothesis, we used the same setup to measure the frequency noise PSD of a reference laser with known phase noise properties (see supplementary material). 
Both lasers had a very similar low frequency behavior, indicating that the DSH method is still limited by acoustical and vibrational pickup of the delay fiber. 
Therefore, the true laser linewidth is likely lower than reported in Fig. \ref{fig:results}.

\section{Conclusion}
In this work, a fully fiber-coupled external cavity DFB-laser setup with intra-cavity EOM-based active stabilization is presented. 
To benchmark the laser performance, its phase noise for different operational states is measured using a short-delay self-heterodyne setup. 
In addition, a digital signal processing method is presented to compensate for the self-interference effect in self-referenced phase noise measurements.
The data-driven method can effectively equalize the phase noise PSDs and prevent interference fringes in the spectrum. 
Using these PSDs, the linewidth of the laser was calculated and shown to be on the order of few \si{\kilo\hertz}.
Both, the laser system as well as the phase noise measurement method described in this work present economical solutions for the problems they are meant to solve for most wavelength bands with off-the-shelf components. 
The \qty{}{\kilo\hertz} linewidth laser enables us to efficiently address  narrow optical atomic transitions for creating an ultracold gas of \isotope[40]{Ca} in a long-lived metastable triplet state. 
Due to its fully fiber-coupled design the laser itself is practically maintenance free.

\section*{Acknowledgments}
This work has been funded by the SPOC Center (Grant No. DNRF 123) and by the Deutsche Forschungsgemeinschaft (DFG, German Research Foundation) (Grant No. HE 2334/15-2). 
The authors also thank Ortwin Hellmig and Hellmtech for help with custom optical coatings and optical fiber splicing.
\section*{Disclosures}
The authors have no conflicts of interest to disclose.
\section*{Data availability statement}
The data and code that support the findings of this study are available from the corresponding author upon reasonable request.

\section*{Supplementary material}
See supplementary material for the estimation and investigation of acoustic noise in the experimental setup.

\bibliographystyle{unsrt}
\bibliography{main}

\section*{\centering Fiber-coupled external cavity DFB-Laser: \\ supplementary material}
\section*{Limitation by acoustic noise}
To test whether our self-heterodyne setup is limited by acoustic noise, we used the setup to measure the frequency noise of a known reference laser. A \texttt{NKT Koheras BASIK E15} at \SI{1550}{\nano\meter} wavelength was measured and processed in the same way as described in the main text. In addition, a separate beat note measurement with two identical copies of the \SI{1550}{\nano\meter} laser in a different lab was performed to acquire a reference frequency noise measurement. As the reference measurement did not feature long delay fibers, acoustic noise should not be a concern.
\begin{figure}[h]
    \centering
    \includegraphics[width=1.0\textwidth]{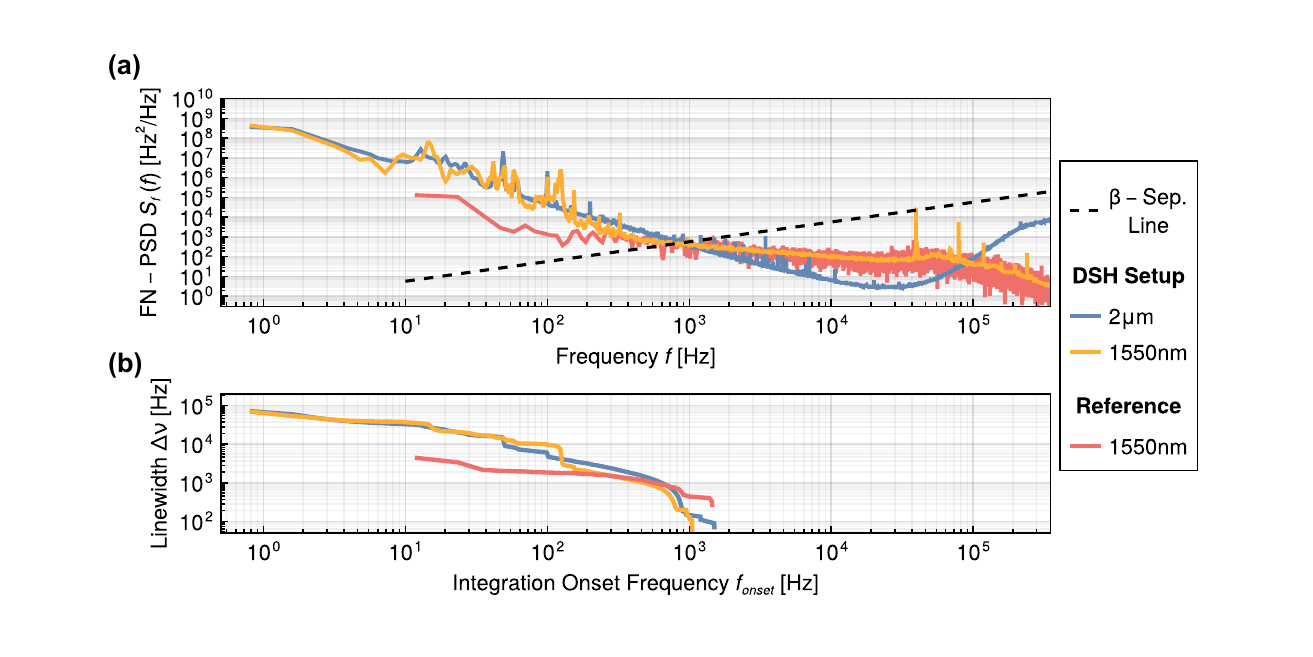}
    \caption{Investigation of acoustic noise in the delayed self-heterodyne setup. \textbf{(a)} Frequency noise power spectral densities of the \SI{2}{\micro\meter} laser under test and a \SI{1550}{\nano\meter} reference laser. The reference laser PSD is measured with the SHD setup and with an independent heterodyne measurement. \textbf{(b)} Corresponding linewidths as calculated via the $\beta$ separation line method.}
    \label{fig:acoustic_noise}
\end{figure}
Both noise spectrum measurements of the \SI{1550}{\nano\meter} laser as well as a representative noise spectrum of the \SI{2}{\micro\meter} laser are shown in Fig. \ref{fig:acoustic_noise}a.
The measurements taken in the DSH setup share the same frequency dependence below \SI{1}{\kilo\hertz}, as well as some of the spurious frequencies.
The spectra of the \SI{1550}{\nano\meter} laser taken with the two different methods overlap very well beyond \SI{1}{\kilo\hertz} as expected. Below however, the DSH measurement displays significantly more frequency noise compared to the reference measurement.
This discrepancy is also visible in the linewidths in Fig. \ref{fig:acoustic_noise}b, where the reference measurement displays an order of magnitude lower linewidths.

This gives a strong indication that for low frequencies the DSH method is still limited by acoustical and vibrational pickup of the delay fiber. Therefore, the true linewidth of the \SI{2}{\micro\meter} laser is likely lower than reported.

\end{document}